\documentclass[conference]{IEEEtran}
\IEEEoverridecommandlockouts
\usepackage{cite}
\usepackage{amsmath,amssymb,amsfonts}
\usepackage{algorithmic}
\usepackage{graphicx}
\usepackage{url}
\usepackage{textcomp}
\usepackage{todonotes}
\usepackage{xcolor}
\usepackage{hyperref}
\usepackage{svg}
\def\BibTeX{{\rm B\kern-.05em{\sc i\kern-.025em b}\kern-.08em
    T\kern-.1667em\lower.7ex\hbox{E}\kern-.125emX}}
\begin{document}

\title{Applying Model-based Requirements Engineering in Three Large European Collaborative Projects: An Experience Report
\thanks{This work has received funding from the ECSEL Joint Undertaking under grant agreements No. 737494 (MegaM@Rt2 project) and No. 101007350 (AIDOaRt project), from ITEA3 (REVaMP2 project No. 15010) and from H2020 under grant agreements No. 732064 (DataBio project) and No. 957212 (VeriDevOps project).}
}

\author{\IEEEauthorblockN{Andrey Sadovykh}
\IEEEauthorblockA{
\textit{Innopolis University} - \textit{SOFTEAM}\\
Innopolis, Russia - Paris, France\\
a.sadovykh@innopolis.ru,\\ andrey.sadovykh@softeam.fr}
\and
\IEEEauthorblockN{Dragos Truscan}
\IEEEauthorblockA{\textit{Åbo Akademi University}\\ 
Turku,Finland\\
dragos.truscan@abo.fi}
\and
\IEEEauthorblockN{Hugo Bruneliere}
\IEEEauthorblockA{\textit{IMT Atlantique, LS2N (UMR CNRS 6004)} \\
Nantes, France \\
hugo.bruneliere@imt-atlantique.fr}
}
\maketitle

\thispagestyle{plain}
\pagestyle{plain}

\begin{abstract}
In this paper, we report on our 5-year's practical experience of designing, developing and then deploying a Model-based Requirements Engineering (MBRE) approach and language in the context of three different large European collaborative projects providing complex software solutions. Based on data collected both during projects execution and via a survey realized afterwards, we intend to show that such an approach can bring interesting benefits in terms of scalability (e.g. large number of handled requirements), heterogeneity (e.g. partners with different types of RE background), traceability (e.g. from the requirements to the software components), automation (e.g. requirement documentation generation), usefulness or usability. To illustrate our contribution, we exemplify the application of our MBRE approach and language with concrete elements coming from one of these European research projects. We also discuss further the general benefits and current limitations of using this MBRE approach and corresponding language.
\end{abstract}

\begin{IEEEkeywords}
Requirements Engineering, Model-based Engineering, Collaborative Projects, Experience Report, Scalability, Heterogeneity, Traceability, Automation
\end{IEEEkeywords}

\section{Introduction}
\label{introduction}
In many countries, funded collaborative projects involving academic partners and industrial ones, either SMEs or large enterprises, are a preferential way of implementing ambitious Research and Innovation Actions (RIAs)~\cite{Van_Noorden2019-bv,European_Commission-DG_CONNECT_undated-ux}. Such projects are also frequently used in order to foster international collaborations and to develop corresponding long-term partnerships between organizations from various countries. This is notably the case in Europe, where the European Commission has several active funding bodies proposing various kinds of funding programmes targeting different societal, economical and scientific grand challenges~\cite{European_Commission_undated-vl}.

In the context of such collaborative projects in the Software and Systems Engineering area, possibly important in terms of the number and variety of involved partners and countries, the main expected results are generally large and complex integrated frameworks or tool sets. In order to allow for their actual design, development and deployment, it is thus fundamental to be able to support and manage as efficiently as possible the corresponding Requirements Engineering (RE) processes~\cite{Ieeeisoiec_undated-ry}, from the initial identification of the industrial needs to their actual realization within the final innovative solutions.

As it can be observed in the literature, the state-of-art in RE is already rich in terms of approaches and corresponding technical solutions~\cite{Cheng2007-lq}. Among other paradigms, some of these approaches rely on model-based concepts \cite{Assar2014-ti}. This notably reflects the fact that model-based principles and techniques have become more popular in industry over the last two decades, as they can provide relevant abstraction, genericity or reusability capabilities (for example)~\cite{Brambilla2012-he}. 

In this paper, we report on our practical experience of proposing and applying a Model-based Requirements Engineering (MBRE) approach and language during 5 years in the context of three different large European collaborative projects, each one of them providing as a result various complex software solutions (e.g. frameworks, integrated tool sets, etc.). The approach was mainly developed by the SOFTEAM company and applied in 3 different research projects in which SOFTEAM participated in the technical coordination activities. With this reporting and data analysis work, we notably intend to show that such a MBRE approach can bring interesting benefits in terms of scalability (e.g. large number of handled requirements), heterogeneity (e.g. partners with different profiles and types of RE background), traceability (e.g. from the initial requirements to the software components), automation (e.g. requirement documentation generation), as well as general usefulness or usability.

The remainder of the paper is structured as follows. Section \ref{contextbackground} introduces the general context and background of this experience report. Then, Section \ref{relatedwork} highlights relevant related work while Section \ref{approach} describes the MBRE approach we proposed in order to support RE processes in collaborative projects. Section \ref{evaluation} evaluates this approach and its practical application in the context of three different large European collaborative projects. Finally, Section \ref{discussion} discusses the main lessons learned from this experience while Section \ref{conclusion} concludes the paper.

\section{Context and Background}
\label{contextbackground}
European collaborative projects are one of the main sources of innovation in Europe~\cite{European_Commission-DG_CONNECT_undated-ux}. There are several funding bodies such as ECSEL, Horizon 2020, ITEA, etc. coming with different types of funding programmes. The funded projects all have in common that a number of organizations (academic or industrial) from several European countries participate in a collaborative research effort. The average number of partners in such projects varies according to the funding framework. For example,  this goes from 4.69 in Horizon 2020~\cite{Accelopment2019-rx} to 30 and 40 for Research and Innovation Actions and Innovation Actions ECSEL projects respectively~\cite{noauthor_undated-ea}. However, it is not unusual that ECSEL projects exceed 100 organizations~\cite{Ecsel-ju_undated-ie}. 

A key element of such European research projects is the complementarity of the project's participants. These participants can come from different application domains (e.g. railway, avionics, telecom, manufacturing), have different company sizes (e.g. from small and medium enterprises to large industrial groups) and levels of maturity related to the project's challenges, or come with different kinds of research background.
They have to work together to achieve a set of shared R\&D goals, usually validated via several case studies that serve as a common platform for experimenting on newly designed and developed technologies. The overall objective is to provide evidence to the European Commission of the benefits and drawbacks, both scientific and economical, that the innovative technologies developed in the context of the project can offer.

However, the diversity and number of partners can also imply project management challenges related to 1) the elicitation of the needs from the industrial case study providers and 2) the identification of not only the concrete solutions to be provided during the project (lasting typically three years), but also of a roadmap for the development of the final technical solution. These challenges are amplified by the number of partners in the project and by the diversity of their (scientific and technical) backgrounds and of their application domains. When challenges are not addressed properly, they can negatively influence the outcomes of the project~\cite{Nepelski2018-gs}. 

Therefore, appropriate RE practices become a necessity for maximizing benefits for a larger number of partners in order to achieve good technical results in such projects. Those RE practices span over all areas of the RE process such as requirements elicitation, specification, validation and management~\cite{Ieeeisoiec_undated-ry}. 
The approach proposed in this paper applies model-based principles and techniques in order to support and improve the RE process in the context of possibly large and diverse collaborative projects. With our approach, we intended to provide better scalability, improved support for heterogeneous types of partners, as well as more complete tool support and automation for managing the requirements of the developed solutions.

In practice, the proposed approach was designed, developed, matured and actually applied in three large European projects covering a period of 5 years (cf. Section \ref{data} for more details and data on these different projects): 
\begin{itemize}
 \item H2020 \textbf{DataBio} 2017-2019 - Data-Driven Bioeconomy \cite{Sadovykh2020-xf} (\url{https://www.databio.eu/}), 
 \item ITEA3 \textbf{REVaMP2} 2016-2019 - Round-trip Engineering and Variability Management Platform and Process \cite{Sadovykh2019-ok} (\url{http://www.revamp2-project.eu/}),
\item ECSEL \textbf{MegaM@Rt2} 2017-2020 MegaModeling at RunTime \cite{Afzal2018-fc} (\url{https://megamart2-ecsel.eu/}). 
\end{itemize}

In this paper, we specifically focus on the three above-mentioned projects as the most recent ones from our side. However, we have already faced the similar challenge to coordinate RE processes and tool framework developments in more than 15 similar European collaborative research projects.

\section{Related Work}
\label{relatedwork}
Requirements Engineering (RE) is the process of identifying, describing, using and maintaining requirements within engineering processes~\cite{Ieeeisoiec_undated-ry,dick2017requirements}. There is a general consensus on the main steps of RE processes~\cite{Pohl2010-xv,van2009requirements}: 1) requirements elicitation, 2) requirements analysis, 3) requirements specification and 4) requirements validation. Requirement management is a transverse activity covering all these steps and ensuring their smooth running. Even if requirements have been used for a long time in different disciplines, the RE advent mostly coincides with the software expansion in the 90's~\cite{kotonya1998requirements}. This resulted in the progressive growth of an active research community working on various related challenges~\cite{nuseibeh2000requirements,Cheng2007-lq}.

By nature, RE always implies a modeling-related activity~\cite{Van_Lamsweerde2000-hj}. Thus, model-based principles and techniques have already been applied in the past to address different kinds of RE activities~\cite{Assar2014-ti}. Notably, this resulted in acknowledged contributions such as goal modeling languages for instance~\cite{yu1997towards,van2001goal} or in the quite recently proposed ReqIF requirement modeling standard~\cite{reqifhomepage}. Besides, RE has also been sometimes applied in the context of model-based development processes~\cite{Loniewski2010-xw}. However, up to our current knowledge, few model-based approaches have been proposed to cover complete RE processes in the general case. For example, an existing solution relies on a generic modeling framework to represent (and simulate) requirements independently from the context~\cite{Baudry2007-nr}. Another one is based on a core requirements metamodel to be customized in order to support various kinds of RE processes~\cite{Goknil2008-jl}. In addition, we can also mention transformation-based approaches from goals or requirements models to design models, such as the KAOS method~\cite{Letier2008-hz}.

Closer to the context and background of our work, a few model-based approaches have already been proposed and used inside collaborative research projects to handle, at least partially, corresponding RE processes. However, either they have been deployed only in the context of a single project~\cite{Karg2016-oz} or they were mostly focusing on some particular RE aspects such as requirements visualization~\cite{Solheim2005-nw}, for instance. In this paper, we rather intend to propose and use a model-based RE approach that is potentially covering more complete RE processes, including important features such as requirements traceability and automated documentation generation. We intend to make such a support practically available in the context of different large collaborative projects involving both heterogeneous and numerous partners.

From a more industrial perspective, and in order to foster genericity, reusability and interoperability, the proposed MBRE approach and language are rooted in well-known and acknowledged software and system modeling standards: UML~\cite{uml} and SysML~\cite{sysml}. Moreover, they are also partially inspired by the European Space Agency standard terminology and structure for RE documents ~\cite{European_Cooperation_for_Space_Standardization2009-jc}.

\section{Proposed Approach}
\label{approach}
As a solution to the problem of covering complete RE processes in the context of large collaborative projects involving many different kinds of partners, we propose a MBRE approach accompanied by a dedicated modeling language and the corresponding tool support. We give an overview of this approach in Section \ref{overviewapproach} and we describe the corresponding RE language we designed in Section \ref{language}. In Section \ref{implementation}, we also provide more details on how these approach and language have been concretely implemented and supported in the Modelio tool.

\subsection{Overview of our Conceptual Approach}
\label{overviewapproach}
Our MBRE approach is natively defined to take into account the typical participant roles in collaborative research projects: \textit{Case Study Providers}  have a practical problem, e.g. in terms of development processes, product quality or features, for which they are looking for innovative solutions; \textit{Research Partners} develop new research methods and prototypes and \textit{Technology Providers} offer technological solutions, in collaboration with the research partners, that can be deployed and evaluated against the industrial case studies. It is the task of the technical coordination team to both ensure a smooth collaboration between the involved parties and make sure that the project's objectives are achieved.

\begin{figure*}[ht]
    \centering
    \includegraphics[width=1.0\textwidth]{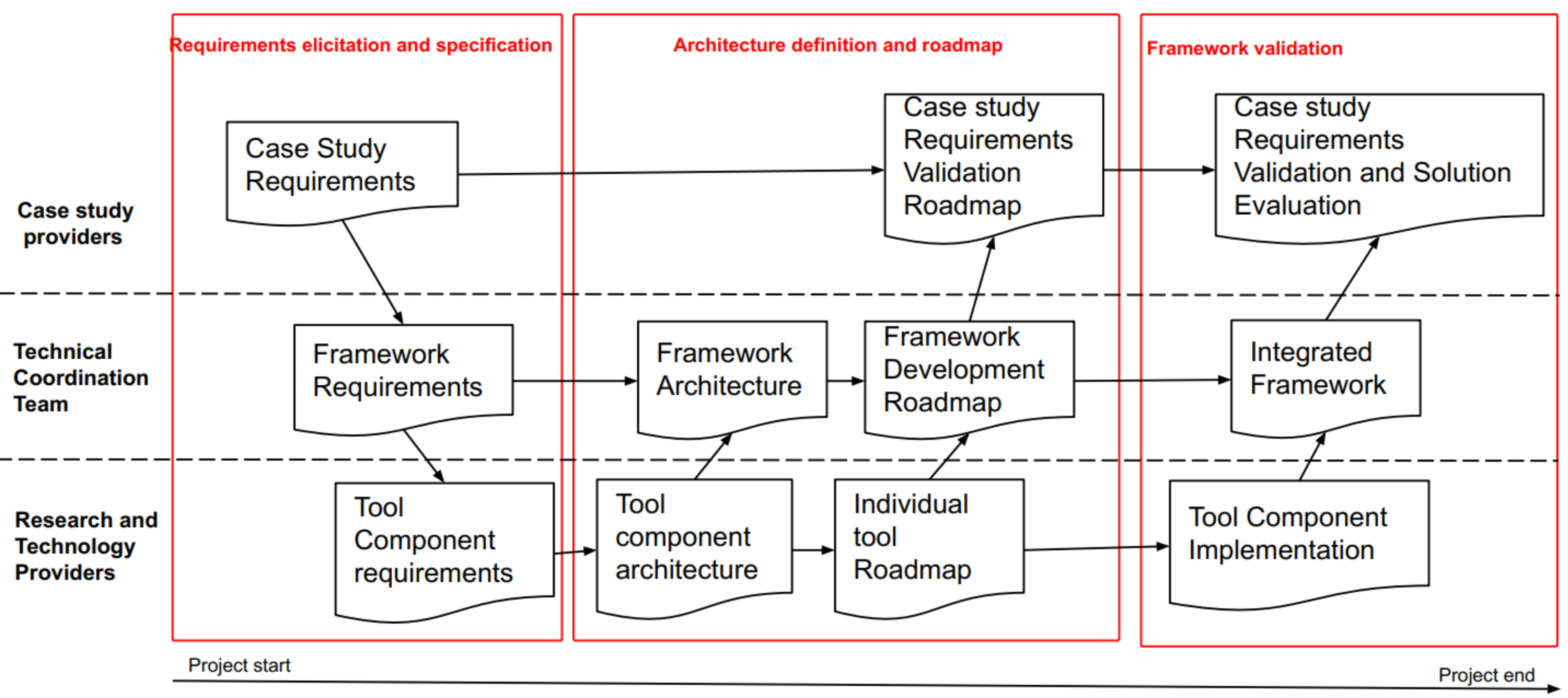}
    \caption{A conceptual approach for MBRE.}
    \label{fig:conceptual}
\end{figure*}

As illustrated in Figure \ref{fig:conceptual}, our approach proposes that the technological solutions in such collaborative research projects are provided in the form of a generic \textbf{framework}. The requirements of the framework are elicited from the requirements of the industrial case studies in the project. The framework aggregates \textbf{Tool Components} which are specific to different application domains and applicable to one or several case studies. Each tool component is developed by different project participants based on \textbf{Tool Component Requirements}, also sometimes called “Tool Purposes", which satisfy the framework requirements. Each tool component will have a specific architecture including the interfaces available for being interconnected with other tools. This notably allows for different tool chains to be easily created for specific case studies. Each tool component is developed throughout the project with different priorities and becomes available at different milestones according to the \textbf{Tool Component Roadmap}. 
	
Based on the individual tool component architectures and roadmaps, the technical coordination team can design the \textbf{Framework Architecture} and the \textbf{Framework Development Roadmap}. Such a roadmap will allow the case study provider to know when different technologies will be available for evaluation, and consequently to create \textbf{Case Study Requirements Validation Roadmaps}.
When different \textbf{Tool Component Implementations} become available at different milestones of the project, they are integrated in the framework and evaluated against the case studies in the \textbf{Case Study Requirements Validation and Solution Evaluation} process.

The above process is applied continuously during the implementation of the project. However, several challenges have to be addressed for ensuring a smooth process. 

The first challenge that we address is the elicitation of the framework requirements that support the needs of the industrial case study providers (i.e., the case study requirements). This step is not trivial since it requires the collaboration of all the partners in the project, each one coming from different application domains and having different technical backgrounds. Having a centralized and clear mapping between case study requirements on one hand, and the framework and tool components requirements on the other hand, also allows the technical coordination team to track the progress of the project, to spot further needs for technical solutions and to mitigate the risks of the project.

The second challenge that we address is to create a roadmap for the development of the framework by collecting development plans for individual tool components. This will allow all partners in the project to be aware of when different features of the framework will be implemented. This will also allow case study providers to know when these tools can be evaluated against their own case studies. In addition, having such a roadmap allows the technical coordination team to better plan and produce different deliverables, demonstrations and thematic events in the project.

\subsection{A Dedicated Modeling Language for RE}
\label{language}

In order to realize the proposed conceptual approach, and to represent and share appropriately the requirement data during the full RE process, we worked on a dedicated modeling language for MBRE. The reason for developing a dedicated language, rather than directly using an already existing one, is that the commonly used modeling languages (i.e. general-purpose ones such as UML) are very wide in terms of scope. Thus, users tend to have multiple and different ways to specify requirements and corresponding design decisions when using them. This is actually an issue in our context of large collaborative projects involving partners with different backgrounds and experiences.

As a solution, we decided to design and build our modeling language by following a bottom-up approach: We started by analyzing what were the documents (i.e. deliverables) generally needed in terms of both requirements and architecture in the context of large collaborative projects. To this end, we notably studied standard representation formats such as ESA ESS~\cite{European_Cooperation_for_Space_Standardization2009-jc} as well as standard modeling languages such as UML~\cite{uml}. Then, inspired by such standards, we designed a generic modeling language that would help in supporting and simplifying the automated generation of these documents/deliverables. The different aspects of the resulting modeling language are presented in what follows.

\subsubsection{Abstract Syntax}
\label{abstractsyntax}
\begin{figure*}[ht]
    \centering
    \includegraphics[width=\textwidth]{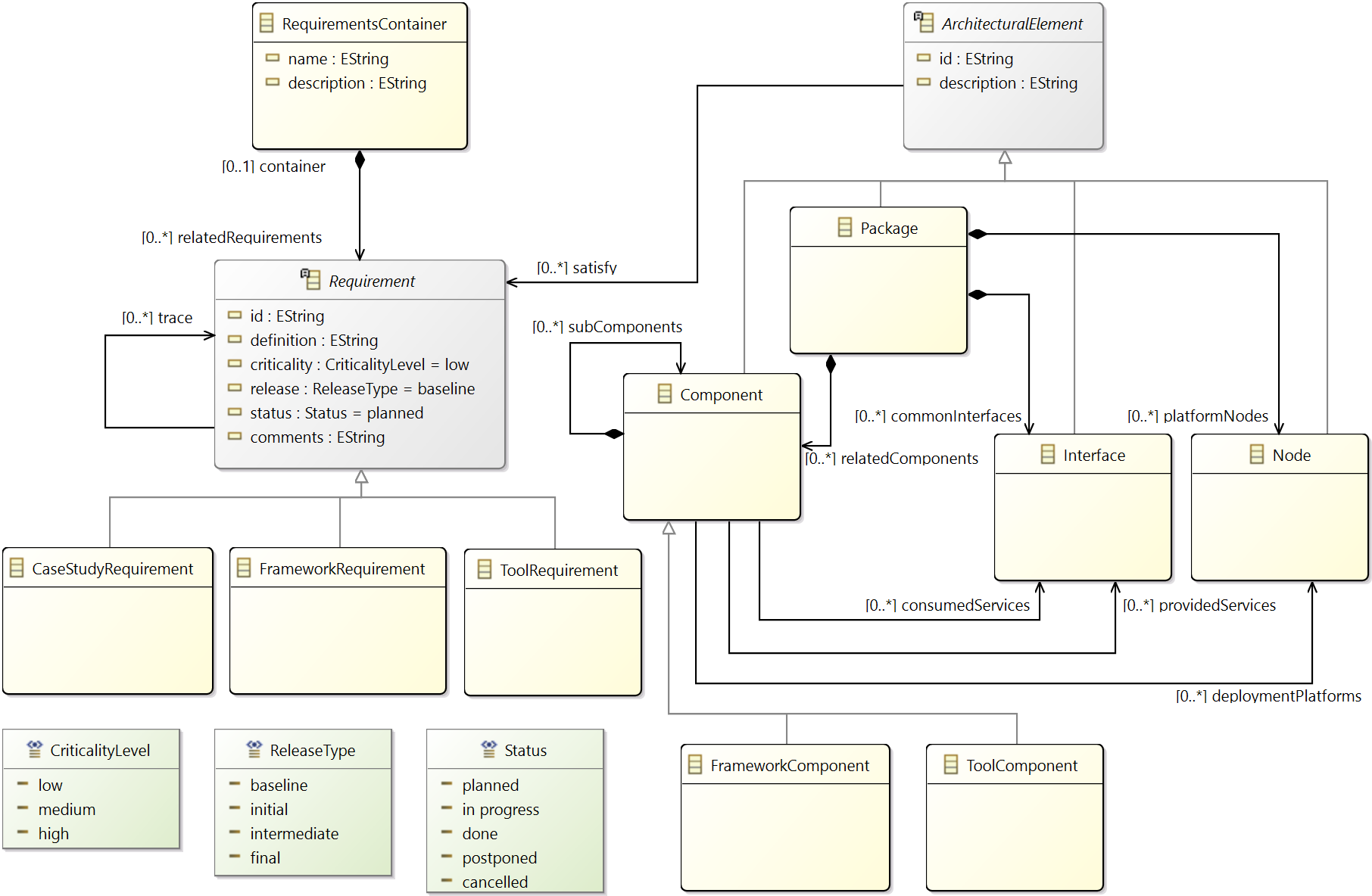}
    \caption{Metamodel describing the abstract syntax of our RE modeling language.}
    \label{fig:metamodel}
\end{figure*}

The main concepts of our dedicated modeling language for RE are depicted in Figure \ref{fig:metamodel}.

At the Requirements level, the base element is the \textbf{Requirements Container} that is used to logically group sets of related requirements (e.g. requirements attached to a same tool). Each \textbf{Requirement} is represented by various properties such as an \textit{identifier}, a \textit{definition}, a \textit{criticality} level, a corresponding \textit{release}, a \textit{status} and additional \textit{comments} if any. The \textit{criticality} level indicates the priority for the requirement’s implementation (i.e. low, medium or high). The \textit{release} indicates the milestone at which the requirement is planned to be satisfied (i.e. baseline, initial, intermediate of final). The \textit{status} indicates the state of fulfillment of the concerned requirement (i.e. planned, in progress, done, postponed or cancelled). The \textit{comments} provide additional information on the concerned requirement. A given requirement can be connected to another requirement it depends on via a \textit{trace} dependency link. For example, a \textbf{Tool Requirement} (also called Tool Purpose) can be linked to a corresponding \textbf{Framework Requirement}, and this Framework Requirement to a corresponding \textbf{Case Study Requirement}. Moreover, any architectural element realizing a given Requirement can be mapped to it via a \textit{satisfy} dependency link.

At the Architecture level, the base element is the \textbf{Package} that is used to logically group sets of related \textbf{Framework or Tool Components}, common interfaces of platform nodes. Each \textbf{Component} depicts a tool or its constituent part. A given component can be composed of different \textit{sub-components} to represent its various constituent parts. An \textbf{Interface} described a required functional service (e.g. XMI import/export). It can be connected to a component via an interface realization link to indicate a \textit{provided service}, e.g. the tool component’s output, or via an use dependency link to indicate a \textit{consumed service}, e.g. the tool component’s input. A \textbf{Node} represents a deployment platform (e.g. Eclipse RCP, a Java virtual machine, etc.). A given component can be connected to different nodes, i.e. its \textit{deployment platforms}. 

\subsubsection{Concrete Syntax}
\label{concretesyntax}
Attached to the abstract syntax of our language as introduced just before, various kinds of concrete notations could be envisioned, either graphical, textual or combining both. Based on our own experiences, and notably the industrial experience of SOFTEAM when using Modelio-based solutions for their customers’ projects, we decided to consider a combination of tabular views and UML/SysML-like diagrams as the syntax for our RE language. 

In terms of diagrams, a SysML Requirements diagram is used to graphically map the different tool components to their respective tool requirements/purposes and then to their corresponding case study requirements. A UML Class diagram is used to both describe a component and their constituent parts and describe its required and provided interfaces. It is also used to map individual tool components to conceptual frameworks components. Finally, a UML Deployment diagram is used to represent the nodes showing the deployment constraints of the different tool components. 

As far as tabular views are concerned, they are mostly used in order to easily enter and then properly display various properties of possibly very large sets of requirements.  

\begin{figure*}[ht]
    \centering
    \frame{\includegraphics[width=\textwidth]{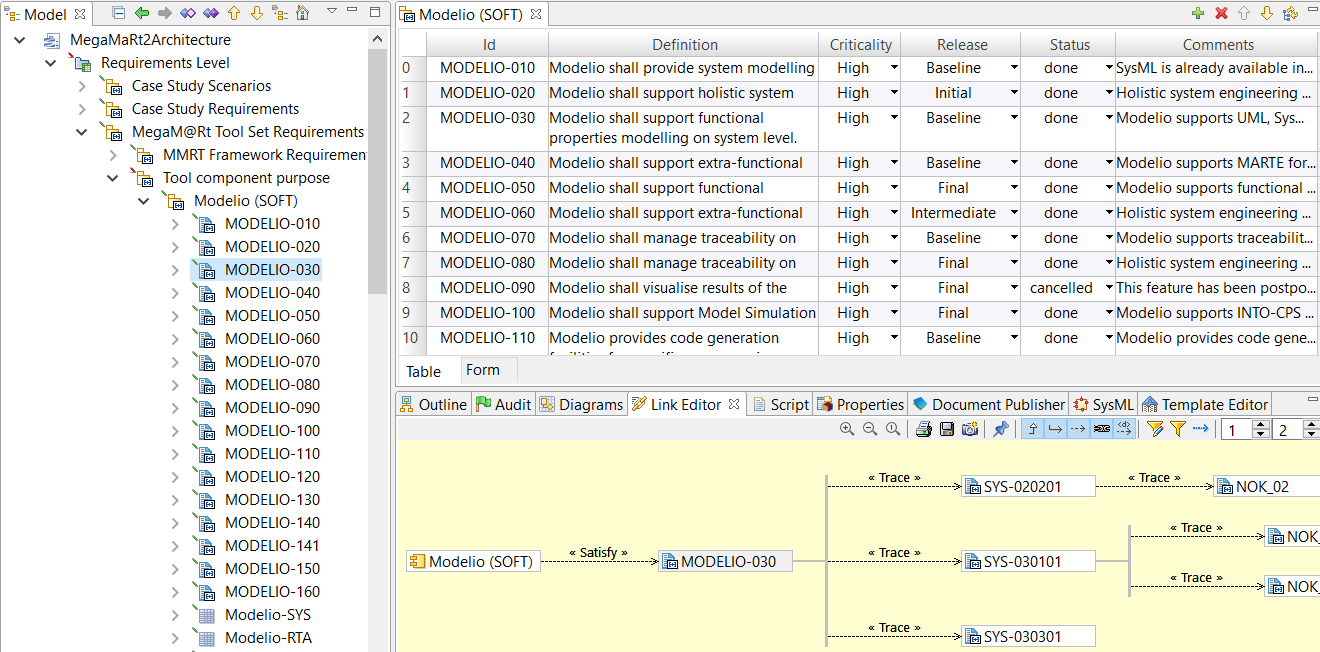}}
    \caption{Example view of the concrete syntax of our RE modeling language, as implemented in the Modelio tool.}
    \label{fig:syntax}
\end{figure*}

Figure \ref{fig:syntax} illustrates this concrete syntax, as currently implemented in the Modelio tool (cf. Section \ref{implementation}), on an example RE model from the MegaM@Rt2 project. In this particular example, we can see in the tabular view a set of tool requirements/purposes for the Modelio tool itself. These tool requirements are grouped in a same requirements container named “Modelio (SOFT)” (cf. left panel). On the related diagram (cf. bottom view), we can see how a given Modelio tool requirement “MODELIO-030” is traced to some framework requirements “SYS-020201”, etc. and then, by transitivity, to some case study requirements “NOK-02”, etc. We can also see that a given tool component “Modelio (SOFT)” satisfies a corresponding tool requirement “MODELIO-030”.

\subsubsection{Semantics}
\label{semantics}
As partially extending UML core concepts, the semantics of our RE language is directly connected to the semantics of UML. For instance, we consider the concepts of \textbf{Package}, \textbf{Component}, \textbf{Interface}, \textbf{Node} and various types of links as having the same meaning that prescribed by UML. However, in order to make the models universally understandable and reusable by other persons relying on our RE language, we limited their usage to the strict definitions provided in Section \ref{abstractsyntax}. 
As introduced earlier, one of the main objectives of our RE language is to allow for the automated generation of corresponding requirements, architecture and roadmap documents. Thus, RE models  expressed in our language are meant to be interpreted by (i.e. taken as inputs of) specific document generators. As a consequence, part of the semantics of our RE language is also embedded in the source code of these document generators. For example, a \textbf{Package} would often correspond to a first-level chapter in the specifications. A \textbf{Common} \textbf{Interface} is a section of the Software Requirements Specification (SRS) where we list the information that is very helpful for integration. Besides, each \textbf{Tool Component} section will have exactly four subsections based on the used diagrams (cf. Section \ref{concretesyntax}): The SysML Requirements diagram is the main source to extract information about the features (existing or planned) or a component, the UML Class diagram maps to the corresponding section giving more information on provided and required services, etc. Basically, the document generator would navigate the whole model from its root and, when getting to an element with a corresponding type e.g. a Package providing Common Interfaces, it will collect the corresponding data from the model and then integrate them into paragraphs, tables, matrices and figures of the generated document, cf. a deliverable generated in MegaM@Rt2 for example\cite{noauthor_undated-kq}.

\subsection{A Modelio-based Implementation of the RE Approach and Language}
\label{implementation}

The complete MBRE approach and language presented in this paper have been implemented in the Modelio tool~\cite{modelio} developed by SOFTEAM. In the DataBio project, we first experimented with extending/refining ArchiMate as the high-level representation for architecture. However, due to the constraints imposed by the ArchiMate metamodel, collaborative modeling was quite tedious in Modelio. For example, the locking of a single element was leading to the locking of a large portion of the model, possibly impacting other users. Moreover, ArchiMate is quite new and much less adopted in the industry. This would have resulted in a steeper learning curve, such a limitation being partially reflected in the survey data presented in Section \ref{survey}. These are the reasons why, in the subsequent REVAMP and MegaM@Rt2 projects, we rather opted for extending/refining UML as 1) allowing atomic editing of components and 2) being already widely known in the industry.

\begin{figure}[ht]
    \centering
    \includegraphics[width=0.7\columnwidth]{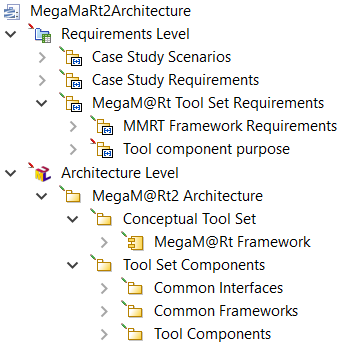}
    \caption{Structure of the RE model in Modelio: The practical example of the MegaM@Rt2 project.}
    \label{fig:REmodel}
\end{figure}

The main elements of a RE model in Modelio are shown in Figure \ref{fig:REmodel}, with respect to the MegaM@Rt2 project as a practical example.
Conforming to our RE language as described in \ref{language}, the overall structure of the RE model includes both the Requirements and Architecture levels. For the Requirements level, we directly benefited from Modelio Analyst features such as Requirements Containers, Tabular, Diagram view, Traceability matrices and  Import from Excel. Modelio also allowed us to specialize the Requirements properties (that correspond to the columns in the tabular view) in order to implement our approach and language. For the Architecture level, we relied on the standard implementation of UML as currently available in Modelio. We refined this UML implementation in order to limit the use of UML to the concepts considered in our RE language. Moreover, in order to facilitate the initial use of the RE language, we also built-in a specific template for tool components and made it available directly from the Modelio workbench. This way, the users of our RE language in Modelio can benefit from clear guidelines on what is expected from them when elaborating on their Requirements model.

The Modelio Document Publisher was also a major feature that helped us to construct our MS Word document generator. Indeed, Modelio provides a simplified document template editor with graphical interface that has pre-build functionalities for navigating the model, filtering model elements, extracting textual notes and diagrams, as well as building sections, paragraphs, tables and matrices. These features were particularly useful, since they allowed us to generate important parts of our documents. This is notably the case for roadmaps, where we created tables displaying mappings between case study requirements and tool requirements indicating planned delivery dates. Based on such roadmaps, the case study providers could anticipate and organize both the building of their tool chains and their validation activities. Moreover, the traceability information contained in the RE model was extremely important in order to identify the case study requirements that are not addressed by any of the tool components. This way, we were able to conduct gap analysis from our RE model in order to better plan corrective actions.

In addition to the base implementation of our MBRE approach and corresponding language, our Modelio-based implementation also provides additional features which are particularly relevant in the context of RE processes within large collaborative projects. For example, it is possible to relate and trace our RE models to elements coming from other kinds of models in Modelio (Enterprise Architecture models expressed in ArchiMate~\cite{archimate}, business models in BPMN~\cite{bpmn}, system models in SysML~\cite{sysml}, etc.). It is also possible to import requirements from external tools (e.g. Microsoft Excel) and to integrate them in our RE model. Moreover, complementary to the already provided document generators, new generation capabilities can potentially be added to target other kinds of text-, diagram-, table- and matrix-based representations for the RE information Finally, our Modelio-based implementation directly benefits from the collaborative features provided by the Modelio environment: support for locking / unlocking model elements, commit and update, user, configuration or version management, etc.

\section{Evaluation}
\label{evaluation}
In order to evaluate our MBRE approach and language, we collected various kinds of data associated with the three large collaborative projects considered in the present experience report. We describe quantitative and qualitative data resulting from both the projects’ execution, in Section \ref{data}, and a survey realized after the end of these projects, in Section \ref{survey}. In Section \ref{assessment}, the collected data is then used to assess the relevance of our RE approach and language according to main targeted properties.

\subsection{General Data on Projects}
\label{data}
The proposed MBRE approach and language have been designed, developed and then deployed in the context of three large collaborative projects, financially supported by the European Commission under different R\&D funding programs.

The first project is named DataBio, standing for Data-Driven Bioeconomy (Horizon 2020). It lasted 3 years from 2017 to 2019 and involved 48 partners from 17 countries, for a total budget of € 15M. 27 industrial pilots / use cases in the agriculture, forestry and fishery areas have been considered during the DataBio project.

The second project is named REVaMP2, standing for Round-trip Engineering and Variability Management Platform and Process (EUREKA ITEA3). It lasted 3 and a half years from 2016 to 2019 and involved 27 partners from 5 countries, for a total budget of € 22M. Seven industrial use cases in the cyber-physical systems, electronic systems or tourism areas have been considered during the REVaMP2 project.

The third and last project is named MegaM@Rt2 \cite{Afzal2018-fc}, standing for MegaModeling at RunTime - An scalable model-based framework for continuous development and runtime validation of complex systems (ECSEL). It lasted 3 years from 2017 to 2020 and involved 27 partners from 6 countries, for a total budget of € 16.7M. 9 industrial use cases in the aeronautics, warehousing, automotive, construction, transportation or telecommunication areas have been considered during the MegaM@Rt2 project.

Complementary to these global projects figures, we provide in Table \ref{tab:noStatistics} additional data displaying the level of activity in terms of RE, as registered during these three projects.

\begin{table}[ht]
    \centering
     \caption{Key figures related to the DataBio, REVaMP2 and MegaM@Rt2 projects in terms of RE activity.
}
    \begin{tabular}{|l|r|r|r|}
    \hline
         & \textbf{DataBio} &\textbf{ REVaMP2} & \textbf{MegaM@Rt2} \\\hline\hline
No. of registered  & 55 & 43 & 56\\
users in the Requirements   & & & \\
  model repository      & & & \\\hline
  
Number of contributors & 31 & 24 & 27\\
 to the Requirements model   & & &\\\hline
                        
 No. of commits to       & 958 & 534 & 1322\\
    the Requirements model   & & & \\\hline
    
No. of ,     & 211 & 535 & 458\\
 handled requirements  &  &  & \\
 in the Requirements model     & & & \\\hline

 Total no. of model elements    & =5052 &=2751 & =3902 \\      
   (Requirements level    & 181 & 535 & 458\\
  + Architecture level) & 4871 & 2216 & 3444 \\
 in the Requirements models  & & & \\\hline
 
  Size of the generated        & 61 & 109  & 125 \\
  SRS document from the & & & \\
   Requirements model   & & & \\
   (no. pages)  & & & \\\hline
    \end{tabular}
   
    \label{tab:noStatistics}
\end{table}

The data show that the number of individuals involved in the RE process, whether they are just registered users having a “read-only” profile or actually active contributors having a read-and-write profile, was relatively important and globally similar in the three projects. 

In terms of RE activities, as illustrated in our case by the number of commits on the Requirements model, we however observed a disparity between the three projects even though it stays globally important in all of them. This can be explained by the slightly different size of the three projects e.g. in terms of number of partners or use cases. This can also be partly explained by the nature of the single commits: A given user can frequently do small single commits to the Requirements model while another can rather commit a large number of updates as a single commit.

In terms of the Requirements models themselves, the data highlight the globally high number of handled requirements and related elements in the context of the three projects. We can observe differences between the projects but, as stated before, this can be explained by the specificity of each particular project e.g. the number of partners and the use cases to be covered. These differences are also directly reflected and visible in the number/size of the various project's documents or deliverables generated from the Requirements models in the three projects. 
We also want to note that the initial use of ArchiMate in the DataBio project, as previously mentioned in Section \ref{implementation}, does not have a significant influence since the number of concepts considered for modeling the architecture is exactly the same than in UML. The slightly bigger numbers can be rather explained by the greater size of the DataBio project in terms of involved partners and tools.

\subsection{Survey for Projects' Participants}
\label{survey}
In order to be able to evaluate our MBRE approach and language according to more data, we have also run a complementary survey among the members of the MegaM@Rt2, REVAMP and DataBio projects. The key research hypothesis we want to validate is whether our approach is relevant and helpful in the context of large collaborative research projects.
We started with three quantitative assessment questions:
\begin{itemize}
    \item \textit{Q1: In your opinion, did you find this graphical model-based approach useful in different activities of Requirements Engineering? Followed by the list of main RE activities.}
    \item \textit{Q2: In your opinion, do you see the modeling approach as an improvement compared to other non-modelling (e.g. text-only or table-based) regarding the following aspects? Followed by the list of characteristics that the requirements have to follow, such as correctness and traceability~\cite{Wiegers2013-ic}.}
    \item \textit{Q3: In your opinion, did you find the following Modelio tool features useful in different Requirements Engineering activities? We listed all presumably key features of Modelio that could be considered helpful.}
\end{itemize}

In addition to these quantitative assessment questions, we proposed to answer three qualitative assessment questions:
\begin{itemize}
    \item \textit{Q4: In your opinion, what was the most challenging aspect of the Modelio-based approach?}
    \item \textit{Q5: In your opinion, what was the most challenging aspect of the Modelio-based approach?}
    \item \textit{Q6: In your opinion, which additional Modelio tool features would have been useful for Requirements Engineering in the project?}
\end{itemize}

The potential respondents were all project partners who had an account in the shared model repository and presumably had access to the modelling. We excluded the authors from the survey in order to avoid the subjectivity bias, even though they were primary beneficiaries of the approach as responsible for architecture definition in the MegaM@Rt2 project.

In total, we had 154 individual contact persons: 55 for DataBio, 43 for REVAMP and 56 for MegaM@Rt2. We received 15 complete answers, including 1 person not involved in RE activities. We explain this level of participation by a couple of factors: (1) Few of these contact persons were active contributors to the RE process in these projects, most of them were “readers” or contributed very few elements; (2) It has been at least 1 year since the projects terminated and at least 2 year since the end of the corresponding RE work. While this amount of data cannot be considered statistically representative, we believe the received feedback still provides interesting and relevant complementary insights.

Overall, 76.95\% of respondents would agree or strongly agree that our model-based RE approach implemented in Modelio was useful. For Q1, 90\% would agree that graphical model-based approaches are useful for RE. For Q2, 65\% would agree the modeling approach is better for RE. For Q3, 79.59\% would agree that Modelio was useful for RE. 

In Q1, all respondents (100\%) agree that model-based approaches are useful in Requirements analysis and negotiation. However, only 80\% find that model-based approaches are useful in Requirements validation.

In Q2, the highest agreement (86.67\%) concerns the advantages of a model-based approach regarding Traceability – can be linked to system requirements, designs, code, and tests. The lowest agreement (53.33\%) concerns the benefits of model-based versus non-model-based approaches for dealing with Correctness – accurately states a customer or external need and Clearness – has only one possible meaning.

In Q3, all respondents (100\%) would find Changing/adding dependencies useful in the implementation of our approach. The lowest agreement (57.14\%) concerns the usefulness of Roadmapping (e.g. setting up expected delivery dates and completion stage for designed components). However, this feature was introduced in the Mega@Rt project only, where 80\% (4/5) of the respondents would find it useful.

Moreover, we asked additional questions on the appropriateness of our approach. This notably resulted in the following assessments:
\begin{itemize}
  \item 85.71\%	would find the approach appropriate for the given size and scope of the project.
  \item 92.86\%	would find the tool support useful for guiding the RE process and enforcing project conventions.
  \item 71.43\%	would find the approach easy to learn.
  \item 78.57\%	would find the approach easy to apply.
  \item 78.57\%	would apply a similar approach in future.
\end{itemize}

Finally, we also received some open qualitative feedback. Respondents generally mentioned the difficulty to convince stakeholders to apply the approach as well as difficulties in terms of model synchronization during collaborative editing. More positively, respondents also mention their appreciation in terms of the standardization of the RE process provided by our approach, the improved traceability support or the fast and clean documentation generation.

For the sake of completeness and transparency, the summary of the survey results is publicly available~\cite{noauthor_undated-cf} as well as the fully detailed answers from the participants~\cite{noauthor_undated-sa}.

\subsection{Overall Assessment of our MBRE Approach and Language}
\label{assessment}
Based on all the collected data presented in Section \ref{data} and Section \ref{survey}), we can provide a first assessment of the relevance of our MBRE approach and language according to the main properties introduced in Section \ref{introduction}:

\begin{itemize}
    \item \textbf{Scalability} - The consolidated data extracted from the Requirements model repository in the three projects clearly shows that we have been able to support a significant number of users, including regularly active ones, as well as to handle a relatively large number of requirements and related model elements. Moreover, the final success of the three projects in terms of deliveries (i.e. documents, tools, demonstrators) and the results of the survey also demonstrate the general scalability of our MBRE approach and language, at least for projects going up to the size of the three mentioned ones.
    \item \textbf{Heterogeneity} - The varied characteristics of the three projects (e.g. different partners providing different kinds of tools and technologies, various use cases covering several application domains) show that we have been able to handle a certain level of heterogeneity. Based on our experience in these three projects, and the participants’ feedback we collected via the survey, we can also argue that our MBRE approach and language can be applied similarly in any large collaborating project whose main purpose is to produce integrated software solutions.
    \item \textbf{Traceability} - Looking at the results of the projects participants survey notably, we can state that traceability and the capability to guide and enforce full RE processes from the beginning to the end of collaborative projects has been one of the most acknowledged features of our MBRE approach and language. Even though improvements and/or extensions are still possible, e.g. tooling support for traceability down to the source code, the feedback received up to now already highlights a correct support for traceability.
    \item \textbf{Automation} - The quantitative data collected at the end of the three projects show that we have been able to automatically generate documents of significant sizes from the corresponding Requirements models. These were quite complete requirements, architecture or roadmap documents whose content could be then reused directly to produce the official projects deliverables. The achieved level of automation, though not total, was already quite appreciated by the projects’ participants.
    \item \textbf{Usefulness and Usability} - The fact that our MBRE approach and language have been successfully used in practice in the context of three different collaborative projects already shows a certain level of usability. The qualitative data collected from the participants’ survey also confirm that the users globally found our approach useful during these three collaborative projects. Furthermore, they appreciated in particular the fact that the approach was easily applicable in their respective contexts, i.e. both easy to learn and to apply.
\end{itemize}

\section{Discussion}
\label{discussion}
In what follows, we discuss further the contributions proposed in this paper as well as our own experience in their contexts. We notably present general lessons learned from the three collaborative projects in Section \ref{lessonlearned}. We also describe in Section \ref{threatstovalidity} a few threats to validity we have identified concerning the current evaluation of our MBRE approach and language. Finally, we introduce in Section \ref{nextstepschallenges} some open challenges concerning both our approach and its applications in other contexts in the future. 

\subsection{Lessons Learned}
\label{lessonlearned}
From our global experience of designing and then applying our MBRE approach and language in the context of three different large collaborative European projects, we have been able to extract some general lessons learned we hope to be useful to the RE community as well as more globally to the whole Software Engineering community:

\begin{itemize}
    \item \textbf{Project management} - The received feedback shows that our MBRE approach and language is mostly beneficial from a project management perspective: One of the most valuable user features appeared to be the automation capability. Notably, the possibility to perform more easily gap analysis and obtain a corresponding roadmap, or the possibility to generate long documents directly from the Requirements model, were highly appreciated.
    \item \textbf{Software architecture} - Moreover, the proposed approach and language also appeared to be relevant from a software architecture perspective. Notably, the mostly appreciated features were related to the support for collaborative work, integration with a model repository, etc. The combination of well-known diagrams coming from UML and SysML with tabular views was also perceived as a good way to facilitate the use of our solution, especially for partners already having experiences in modeling or related activities.
    \item \textbf{Learning curve} - Some participants in our projects had a somehow limited previous experience in modeling, both conceptually and technically. This resulted in difficulties for them to catch up with some of the concepts proposed by our RE language. In this same vein, it also appears that the guidelines initially provided, both on the approach/language and on the Modelio tooling itself, were not sufficiently detailed in order to allow for a first usage. Fortunately, this was quickly fixed by the Modelio team thanks to the organization of online hands-on sessions for example. The participants were then more easily able to go on with their RE activities in the context of their respective projects.
\end{itemize}

\subsection{Threats to Validity}
\label{threatstovalidity}
In this paper, we evaluated the proposed MBRE approach and language according to different kinds of data we were able to collect both during and after the end of the three European collaborative projects mentioned earlier.

The main threat to validity probably concerned the amount of data we have been able to gather. Our MBRE approach and language have been deployed in "only" three different projects from which we have extracted mostly quantitative data. However, the fact that these were large projects that run over 5 years in total already provides a certain level of confidence about the quality and relevance of the collected data. Moreover, we complemented this quantitative data with extra data (quantitative but also qualitative this time) we obtained as a result of a survey largely distributed among the three projects participants. Obviously, it would have been more significant to get more answers to the survey. However, we believe the collected feedback coming from 15 different participants is already interesting in order to improve our global appreciation of the proposed solution.

\subsection{Next Steps and Challenges}
\label{nextstepschallenges}
The very next step of our work will consist in improving our experience with the proposed MBRE approach and language by applying in the context of another large collaborative project. This will actually be the case as we are now starting RE activities in the context of two new European projects - VeriDevOps (H2020)~\cite{sadovykhveridevops} and AIDOaRt (H2020 ECSEL) that will run 3 more years and involve more than 40 partners, both industrial and academic ones, coming from 7 different countries. As also strongly involved in these new projects, we (the authors) have already planned to work on deploying again our MBRE approach and language. This way, we hope to be able to 1) collect more relevant feedback from our partners and 2) upgrade our solution accordingly. For instance, similarly to what we did in the MegaM@Rt2 project, we plan to experiment further on the use of our MBRE approach and language in the context of hackathons mixing both business and technical people~\cite{Sadovykh2019-nl}.

Moreover, as a result of the work and overall experience  reported in this paper, we also identified some challenges we believe to be worth investigating as far as MBRE or more generally RE is concerned: 

\begin{itemize}
    \item \textbf{From requirements to source code} - Our requirements models contain information concerning mostly the needs and architectural decisions at the project level. While this is already useful for coordinating the common global effort towards the realization of the target solution, it remains relatively far from being full model-driven development where implementation and verification artifacts can be produced from the models. Thus, more efforts still have to be made in order to better mind this gap between the architecture and development levels in our MBRE solution and also possibly in others.
    \item \textbf{User training and support} - In our RE language, we deliberately restricted the UML usage to a limited number of concepts. We also provided tooling support, user guidelines and online workshops to make the life of the various projects’ partners easier. Nevertheless, the partners appeared to still need constant support with the tooling and the approach. Thus, we strongly believe that the usability and learning curve of RE solutions are key elements to consider and improve accordingly in order to allow for their large industrial dissemination.
    \item \textbf{Collaborative and online work} - One of the most reported issues concerned restrictions in the collaborative editing capabilities provided by our MBRE solution as relying on a Subversion-based lock - edit - commit - release operation mode. The fact is that, nowadays, users generally tend to prefer online editing collaboration modes e.g. via their favorite Web browser. However, we have seen limited support for that so far in existing modelling tools or even in popular IDEs. Such a support for online collaboration at the requirements level is probably a path to be explored more deeply in the future as far as RE solutions are concerned.
    \item \textbf{Automation and production} - There are still open challenges related to the support for automation in RE processes and more generally in Software Engineering processes. For instance, we could have considered to build-in some more automated support for document generation or even code generation. However, it is always a matter of cost/benefit balance since development resources can be limited in collaborative R\&D projects. Thus, our MBRE solution is still not in full production stage e.g. it requires some level of customization for each new project, and it will need more work to be made available as an actual product in the future.
\end{itemize}

\section{Conclusion}
\label{conclusion}
In this paper, we reported on our practical experience of proposing and deploying a Model-based Requirements Engineering approach and language during 5 years in the context of three different large European collaborative projects providing complex software solutions. Our MBRE approach and language mostly focused on supporting three complementary aspects: 1) requirements are described appropriately at different abstraction layers (case study, framework, tool), 2) requirements can be better interconnected and traced during the RE process and 3) requirements can be used to (semi-)automatically perform gap analysis, roadmap and corresponding document generation.

Based on this global experience and the collected data, we showed that our MBRE approach and language can bring interesting benefits in terms of scalability, heterogeneity, traceability, automation, general usefulness or usability. We also discussed the added-value of our solution from a project management and architecture perspective while identifying some limitations we faced, in terms of RE solution learning curve for instance, and that we already managed to partially overcome. As a conclusion, we have concrete plans to continue working on extending and applying our MBRE approach and language in the context of other large collaborative projects we will be involved in the near future.

\newpage


\end{document}